\documentclass[prd,preprint,a4paper,superscriptaddress,nofootinbib,11pt]{revtex4}
\usepackage{amssymb}
\usepackage{amsfonts}
\usepackage{amsmath}
\usepackage{amsfonts}
\usepackage{amssymb}
\usepackage{hyperref}

\def\fs{\footnotesize}

\begin{document}

\title{Dark Energy Dominance and Cosmic Acceleration in First Order
Formalism}
\date{\today}

\author{Gianluca Allemandi}
\email{allemandi@dm.unito.it}
\affiliation{\fs Dipartimento di Matematica,  Universit\`a di Torino\\
Via C. Alberto 10, 10123
Torino (Italy)}
\author{Andrzej Borowiec}
\email{borow@ift.uni.wroc.pl}
\affiliation{\fs Institute of Theoretical Physics, University of
Wroc{\l}aw\\
Pl. Maksa Borna 9, 50-204  Wroc{\l}aw (Poland).}
\author{Mauro Francaviglia}
\email{francaviglia@dm.unito.it}
\affiliation{\fs Dipartimento di Matematica,  Universit\`a di Torino\\
Via C. Alberto 10, 10123
Torino (Italy)}
\author{Sergei D. Odintsov}
\email{odintsov@ieec.uab.es also at TSPU, Tomsk}
\affiliation{\fs  Inst. Catalana de Recerca i Estudis Avancats
(ICREA),
Barcelona  and  Inst. d'Estudis Espacials de Catalunya (IEEC)/ICE(CSIC),
Campus
UAB,
Fac. Ciencias,
Torre C5-Par-2a E-08193 Bellaterra (Barcelona) Spain}

\pacs{98.80.Jk, 04.20.-q}

\begin{abstract}
The current accelerated universe could be produced by modified
gravitational dynamics as it can be seen in particular in its Palatini formulation.
 We analyze here a specific non-linear gravity-scalar system in the first
order
\textit{Palatini} formalism which leads to a FRW cosmology
 different from the purely metric one. It is shown that the emerging FRW cosmology may lead either to
 an effective quintessence phase (cosmic speed-up) or to an effective phantom
phase. Moreover, the already known {\it gravity assisted dark energy dominance} occurs
also in the first order formalism. Finally, it is shown that a 
dynamical theory able to resolve the cosmological constant problem
exists also in this formalism, in close parallel with the standard metric formulation.

\end{abstract}

\maketitle

\section{ Introduction }

The dark energy problem (for astrophysical data indicating dark energy
existence
and related universe acceleration see \cite{Perlmu} and refs. therein)
is
considered to be a challenge for modern Cosmology. There are various
approaches to its construction.
In general, one can classify most efforts  as:
a) modified gravity (for instance, the so-called $f(R)$ theory); or: b) introducing
some
exotic matter with negative pressure. The {\it truth}, as usual, may lie in
between.
Indeed, since the two directions above are completely phenomenological one can go
further and try to modify gravity via a non-standard (non-minimal)
coupling
with matter. This may in fact provide the effective description for a dark energy universe.
The importance of such models has already been pointed out in
\cite{GADED}
where a possible origin for
  dark energy dominance was suggested.  In \cite{muko, dolgov} a suitable non-linear
matter-gravity theory was used to construct the dynamical approach to the 
resolution of the cosmological constant problem.
(Such coupling may be used for cosmological purposes as in
\cite{PBM}).
Here, the gravity-matter coupling part is assumed to be a $f(R)$
theory-like
Langrangian, non-minimally coupled with a scalar field Lagrangian. It is
interesting   that in such models the standard conformal transformation
to
scalar-tensor theory is over-complicated (if possible at all). This implies
that
in the purely metric formalism only one possible frame really exists. A first aim of ours is 
to see how much the picture changes if one allows a first-order \'a la Palatini approach to the gravitational field.\\
\\
In the recent study of modified gravity with a non linear $1/R$ term \cite{carrol2} (for
 its one-loop quantization see \cite{eli}) in relation with cosmic
acceleration it
was further drawn the important lesson that the 
metric formalism is not equivalent to the Palatini (first-order) formalism.
(Actually, this fact was known in general after the early proof given in \cite{FFV}).
Indeed, while modified gravity of only special form may be viable in
 the metric version \cite{metricfR} (due to a possible instability
\cite{do}) its Palatini version may be instead viable (see
\cite{palatinifR,ABFR} and references therein). Of course, the
cosmological
dynamics of both versions are in principle different. Notice, however,
that modified gravity may be always rewritten as a (well-studied) scalar-tensor
theory.\\
It is clear that a generic non-linear matter-gravity theory may lead also
  to non-equivalent gravitational physics, depending on the choice of either
metric or Palatini description.\\ 
In \cite{GADED} it has  been shown how dark energy may become
  dominant over standard matter at small curvatures in the metric formulation.
  This is directly caused by the choice of the Lagrangian.
  Non-linear gravity-scalar field systems and modified gravity may nevertheless provide
  an effective phantom or quintessence dark energy with the possibility of 
explaining the present speed-up of the universe (and without the need of introducing 
negative kinetic energy fields) \cite{GADED}. \\
It has been  proven in \cite{muko}  that a dynamical approach to the
cosmological constant based on a suitable non-linear gravity-scalar field system
with
a scalar field potential having a minimum at a generic, but negative
value provides a possible solution to the cosmological constant problem. This
is
based on the presence of a non-standard kinetic term in the scalar field
Lagrangian, which diverges at zero curvature. In this particular model
the
cosmological constant freezes at a nearly zero value, while field equations are
stable
under radiative corrections and they provide the expected asymptotic
limit. \\

The same class of non-linear matter-gravity theories are here  studied in the
Palatini
formalism (where the purely gravitational part is suitably generalized) and applied to alternative gravity. This formalism,
considering the metric and the connection as independent fields,  produces second
order field equations which are not affected by  instability problems and
 seem to be in a fairly acceptable accordance with   solar system experiments \cite{palatinifR}.
The Palatini version leads to a completely new formulation when 
compared  with the standard metric formulation and it provides cosmological models which
are able to explain also the present time cosmic speed-up.  \\
It is also shown that the Palatini formalism in the case of non-linear
gravity-scalar  systems gives an acceptable realization of the  dark energy dominance and the
resolution of the cosmological constant problem. Thus, even if the metric and the Palatini
formulations apparently  lead  to different gravitational physics,
  one can see that the same cosmological phenomena may qualitatively occur 
in both frameworks, which therefore deserve to be carefully compared.

\section{ First-order non-linear Gravity-Matter systems: \\ the First
Order
formalism }
We consider a $4$-dimensional Lorentzian manifold $M$, with a metric
$g_{\mu
\nu }$ and an
independent connection $\Gamma^\alpha_{\mu \nu }$. The starting
non-linear
gravity-matter   Lagrangian is:
\begin{equation} \label{lagruno}
\ L=\sqrt{g}\left( F(R)+f\left( R\right) L_{d}\right) + \kappa L_{mat}
(\Psi)
\end{equation}
  where  $f(R)$ and $F(R)$ are some analytic functions of the 
scalar field $R=g^{\alpha\beta}R_{\alpha \beta}(\Gamma )$ and $L_{d}$ is a scalar field  Lagrangian; 
here we set $L_{d}=-\frac{1}{2}g^{\mu \nu }\partial _{\mu }\phi $ $\partial_{\nu
}\phi +V \left( \phi \right)= - \frac{1}{2}\parallel  \nabla \phi \parallel^2+V \left( \phi \right) $. In principle,  $L_{d}$ could be any other matter Lagrangian
(describing spinors, vectors, etc). Moreover, one can think of more general cases when
$f(R)$ (or even a function of Ricci squared terms \cite{Ricci2}) couples
to the Riemann (Ricci) tensor \cite{carroll1} contracted with some derivatives of matter and multiplied by a  similar matter
Lagrangian. It is important to remark that in all cases (even if $L_{d}$ is chosen to be
the
Lagrangian of a scalar theory!) the general class of theories described by
phenomenological Lagrangians of the form (\ref{lagruno}) cannot be conformally
transformed
into an equivalent scalar-tensor theory, with the only exception of the theories already treated in
\cite{ABFR}
and \cite{palatinifR} (i.e. all the theories in which $L_d=1$, i.e. $f(R)$ theories). The term \ $L_{mat}$ \ represents a matter Lagrangian,  functionally
depending on arbitrary matter fields $\Psi$ together with their first derivatives,
equipped with a gravitational coupling constant $\kappa=8\pi G$, which
can be
supposed to be $\kappa=8\pi G=1$ in natural units.\\
In the first order \textit{Palatini} formalism  the Ricci-like scalar is
$R\equiv R( g,\Gamma) =g^{\alpha\beta}R_{\alpha \beta}(\Gamma )$ where $
R_{\mu \nu }(\Gamma )$ is the Ricci tensor of any independent
torsionless
connection $\Gamma$; see \cite{ABFR, FFV} for details. Equations of motion are respectively:
\begin{enumerate}
\item \textit{Field equations for the metric field $g$:}
\begin{equation} \label{eqng}
 F^\prime (R) R_{\mu \nu }-\frac{1}{2} F(R) g_{\mu \nu }+f^{\prime }\left( R\right)
L_{d} R_{\mu \nu }-\frac{1}{2}f\left( R\right) L_{d}g_{\mu \nu }+f\left(
R\right) T_{\mu \nu }=\tau _{\mu \nu }
\end{equation}
where $\ T_{\mu \nu }=-\frac{1}{2}\partial _{\mu }\phi $ $\partial _{\nu
}\phi $ \ is stress-energy tensor of the scalar field and $\tau _{\mu
\nu}\equiv T_{\mu\nu}^{mat}$ \ is the matter stress-energy tensor generated by taking the
variational derivative of $L_{mat}(\Psi)$.
Taking the trace of  equation (\ref{eqng}) with respect to
the metric $g^{\mu \nu}$  we obtain the scalar-valued master equation:
\begin{equation} \label{struct}
 F^\prime (R) R- 2  F (R)+\left( f^{\prime }\left( R\right) R-2f\left( R\right) \right)
L_{d}+f\left( R\right) T=\tau
\end{equation}
where  $\tau =g^{\mu \nu }\tau _{\mu \nu }$ and $ T=g^{\mu \nu}T_{\mu
\nu }=L_{d}-V\left( \phi \right)=- \frac{1}{2}\parallel \nabla \phi \parallel^2 $. The above equation can be
re-written, more
conveniently,  as:
\begin{equation*}
2 F(R)-  F^\prime (R) R +\tau =\left( f^{\prime }\left( R\right) R-f\left( R\right) \right)
L_{d}-f\left( R\right) V\left( \phi \right)
\end{equation*}
Hence, the scalar-valued equation (\ref{struct}) controls solutions of
equation
(\ref{eqng}). We shall refer to this scalar-valued equation as the
\textit{structural equation} of spacetime.
For any real solution $R=F(\tau,T)$ of (\ref{struct}) it follows that
$f(R)=f(F(\tau,T))$ and $f^\prime(R)=f^\prime(F(\tau,T))$ can be seen
as
functions of the traces $\tau$ and $T$ of the stress-energy tensors of
matter
and of the scalar field $\phi$. For  convenience we shall use the abuse of notation
$f(\tau,T)=f(F(\tau,T))$ and $f^\prime(\tau,T)=f^\prime(F(\tau,T))$.\\
Thus,  in the same way as in \cite{ABFR, Ricci2}, the generalized
Einstein equations may be introduced:
\begin{equation} \label{eingen1}
R_{\mu \nu }\left( \Gamma \right) =g_{\mu \alpha }P_{\nu }^{\alpha }
\end{equation}
where one defines the operator $P_{\nu }^{\mu }=\frac{c}{b}\delta _{\nu
}^{\mu
}-\frac{f\left(
R\right) }{b}T_{\nu }^{\mu }+\frac{1}{b}\tau _{\nu }^{\mu }$. The scalars $b$ and $c$, depending on $R$, are here defined as:
\begin{equation} \label{netnet}
\begin{cases}
b=b(R)=F^\prime (R)+f^{\prime }\left( R\right) L_{d}\cr
c=c(R)=\frac{1}{2}\left( F(R)+f\left( R\right) L_{d}\right)
=\frac{(L-L_{mat})}{2 } \sqrt{g}
\end{cases}
\end{equation}
(see also Section \ref{sec3}).

\item \textit{Field equations for the connection field $\Gamma$:}
varying  the Lagrangian
(\ref{lagruno}) with respect to the connection one gets:
\begin{equation}
\nabla _{\lambda }^{\Gamma }\left[\sqrt{g}\,g^{\mu \nu }\left( F^\prime (R)+f^{\prime }\left(
R\right) L_{d}\right) \right] =0
\end{equation}
which states that $\Gamma$ can be chosen to be the Levi-Civita
connection of the metric $h_{\mu\nu }=\left(
F^{\prime }\left( R\right)+f^{\prime }\left( R\right) L_{d}\right) g_{\mu \nu }\equiv bg_{\mu
\nu}$, providing the spacetime manifold with the
typical bi-metric structure ensuing from the first-order  Palatini formalism \cite{palatinifR}. This implies that the
generalized Einstein  equation (\ref{eingen1}) can be immediately rewritten as:
\begin{equation} \label{hmeGEE}
R_{\mu \nu }\left( h \right) =g_{\mu \alpha }P_{\nu }^{\alpha }
\end{equation}
where $R_{\mu \nu }(h)$ is now the Ricci tensor of the new metric $h$, being $\Gamma=\Gamma_{LC} (h)$.

\item \textit{Field equations for the scalar (dilaton-like) field $\phi$:}
performing the variation of the Lagrangian (\ref{lagruno}) with respect
to
$\phi$ we obtain:
\begin{equation}
\partial _{\nu }\left( \sqrt{g}f(R)g^{\mu \nu }\partial _{\mu }\phi
\right) =- \sqrt{g}f(R) V^{\prime }\left( \phi \right)
\end{equation}

\item \textit{Field equations for the matter fields $\Psi$:} they can be
obtained from the Lagrangian (\ref{lagruno}) by the usual prescription $\frac{\delta L_{mat}
(\Psi)}{\delta \Psi}=0$, depending on the particular form of $L_{mat}
(\Psi)$.
\end{enumerate}

Before proceeding further let us make two important notes.
Firstly, in order to get Eq.(6) one has to additionally assume that $L_{mat}$ is
functionally independent of an arbitaray connection $\Gamma$; however it may contain
metric covariant derivatives $\nabla^g$ of some matter fields denoted here by $\Psi$. 
This means that the matter stress-energy tensor $T^{mat}_{\mu\nu}\equiv\tau_{\mu\nu}$ 
depends  solely on the metric $g$ and fields  $\Psi$ together with their partial 
derivatives. Thus, usually this tensor is covariantly conserved, in a sense of metric covariant derivative, i.e. $\nabla_g^\nu\tau_{\mu\nu}=0$. Nevertheless, we do not assume
a priori that the (metric) covariant derivative of the left-hand side of Eq. (2) identically vanishes (in fact it does not in general). However this does
not provide any inconsistency in our model. When we solve the field equations of motion 
with a source - here the right-hand side of Eq. (2) - having it a vanishing covariant diveregence thus also the left-hand side of Eq. (2) taken on this particular solution 
(i.e. "on shell") will have automatically vanishing covariant derivative as well.

Secondly, we are now close to discuss energy momentum conservation in modified 
gravity models. As it is well-know that any generally covariant field theory
admit the covariant conservation of energy momentum which can be expressed in 
the form of generalized Bianchi identitiy. This general fact has been presented 
in many books and articles and it is widely recognized as (second) Noether theorem
(see e.g. a recent book  \cite{FF04} and ref.s quoted therein).
More in particular, for the case of modified (including scalar-tensor)
theories of gravity  (also within a Palatini formalism)  the issue has
been  nicely discussed  in a recent paper  \cite{Koivisto}, where explicit 
formulae for the Bianchi identity can be also found.


\section{FRW cosmology from generalized Einstein
equations \label{sec3}}
Owing to the astonishing cosmological experimental results recently obtained in \cite{Perlmu}
 and to the renewed interest in alternative theories of Gravity for cosmological applications,
 it is quite interesting to study the cosmological implications
of the formalism above. Let $g$ be a Friedmann-Robertson-Walker (FRW) metric:
\begin{equation} \label{FRW}
g=-d t^2+a^2 (t) \Big[ {1 \over {1-K r^2}} d r^2+ r^2 \Big( d \theta^2
+\sin^2
(\theta) d \varphi^2  \Big) \Big]
\end{equation}
where $a (t)$ is a scale factor and $K$ is the space curvature
($K=0,1,-1$).
The stress-energy tensor of a matter (ideal fluid) is 
\[
T_{\mu\nu}^{mat}\equiv \tau_{\mu\nu}=
(\rho+p)u_\mu u_\nu+pg_{\mu\nu}
\]
where $p$ is the pressure, $\rho $ is the density of matter and $u^\mu$ is
a co-moving fluid vector, which in a co-moving frame
($u^\mu=(1,0,0,0)$) becomes simply:
\begin{equation}
\tau_{\mu \nu}=
\left(
\begin{array}{clcr}
\rho&0&0&0\\
0&\frac{pa^2 (t)}{1-K r^2}&0&0\\
0&0&pa^2 (t)r^2&0\\
0&0&0&pa^2 (t)r^2\sin^2 (\theta)
\end{array}
\right) \label{Tmunu}
\end{equation}
The standard relations between the pressure $p$, the matter density $\rho$
and the expansion factor $a(t)$ are assumed
\begin{equation}
p=w\rho \quad , \quad
\rho=\eta a^{-3(1+w)} \label{pro}
\end{equation}
where particular values of the parameter $w\in \{-1,0,{1\over 3}\}$ will
correspond to the
vacuum,  dust or radiation   dominated universes. Exotic matter (which is
up to now under investigation as a possible model for dark energy) admits
instead
values of $w<-1$; see e.g. \cite{Perlmu}. These  expressions (\ref{pro}) follow
from the metric covariant conservation law of the energy-momentum  
$\nabla_g^\mu \tau_{\mu \nu}=0$ (as dicussed above) and consequently  
the continuity equation  should hold:
\begin{equation}
\dot{\rho}+3H(\rho+p)=0
\end{equation}
where $H=\frac{\dot a}{a}$ is the \textit{Hubble constant}.\\
We  recall here that the field equations for the metric field $g$ can now be rewritten  in terms
 of the new metric $h$, as already done in  (\ref{hmeGEE}).  Relying on the  methods already developed in \cite{ABFR} it is  simple to see that
the generalized FRW equations following from  (\ref{hmeGEE})  are
\begin{equation}
  \left( \frac{\overset{.}{a}}{a}+\frac{\overset{.}{b}}{2b}\right) ^{2}+
\frac{K}{a^{2}}=\frac{1}{2}P_{1}^{1}-\frac{1}{6}P_{0}^{0}
\end{equation}
where $P_{1}^{1}$ and $ P_{0}^{0}$ are the components of 
  $P_{\nu }^{\mu }=\frac{c}{b}\delta _{\nu }^{\mu }-\frac{f\left(
R\right) }{b}T_{\nu }^{\mu }+\frac{1}{b}\tau _{\nu }^{\mu }$. We recall  here  that the
parameters $b$ and $c$ are  defined (\ref{netnet}). Owing to the cosmological principle and to the consequent homogeneity
and isotropy of spacetime,
we argue that  $\ T_{\mu \nu }$, $\tau _{\mu \nu }$ and $\phi$ depend
just on
the time parameter $t$. As already noticed
in \cite{ABFR, Ricci2} for much simpler cases, we get from the structural equation  (see
 the  above discussion) that both $b$ and $c$ can be expressed just as functions
of time. \\
In our case, we have moreover that $L_{d}=\frac{1}{2}\overset{.}{\phi
}^{2}+V\left( \phi \right) $, so that:
\begin{equation}
\begin{cases}
P_{1}^{1}=\frac{c}{b}-\frac{f\left( R\right)
}{b}T_{1}^{1}+\frac{1}{b}\tau
_{1}^{1}=\frac{c}{b}+\frac{p}{b}\cr
P_{0}^{0}=\frac{c}{b}-\frac{f\left( R\right)
}{b}T_{0}^{0}+\frac{1}{b}\tau
_{0}^{0}=\frac{c}{b}-\frac{f\left( R\right) }{2b}\overset{.}{\phi }^{2}-
\frac{\rho }{b}
\end{cases}
\end{equation}
The modified FRW equation can  be thus written as:
\begin{equation} \label{utrec}
\left( \frac{\overset{.}{a}}{a}+\frac{\overset{.}{b}}{2b}\right)
^{2}+\frac{K%
}{a^{2}}=\frac{1}{6}\frac{f\left( R\right) }{b}\left( L_{d}-V\right)
+\frac{c%
}{3b}+\frac{3w+1}{6b}\eta a^{-3\left( 1+w\right) }
\end{equation}
and the non-minimally coupled scalar-field equation reduces to:
\begin{equation} \label{effLdll}
  \frac{d}{dt}(\sqrt{g}f(R)\overset{.}{\phi
})= \sqrt{g}f(R)\,V^{\prime }\left( \phi \right) 
\end{equation}
 From now on we assume, for simplicity and following the headlines of
 \cite{GADED},  that $V\left( \phi \right) \equiv 0$. This implies that
 the field equations can be written as $\sqrt{g}f(R)\overset{.}{\phi }=const$ or  equivalently  $gf(R)^{2}L_{d}=\beta ^{2}=const$. It follows that, 
on-shell with respect to the above field equations for the scalar field $\phi$, the  
non-minimal coupling term in the Lagrangian takes the form 
\begin{equation} \label{effLd}
f(R)L_{d}=\beta^{2} f(R)^{-1}a^{-6},
\end{equation} 
which differs from the
results of \cite{GADED}  obtained in the metric formalism in the particular case $F(R)=R$, which is in fact degenerate (see below). 


\subsection{A particular case: $F(R)=R$ \label{trippa}}
We start here considering the degenerate case  $F(R)=R$.  This case is worth to be considered as it is the simplest case of non-linear gravity-scalar system reproducing General Relativity in   the limit $f(R)=0$; results can moreover be simply compared with results already obtained  in \cite{GADED} 
in the metric formalism.  We stress that to obtain an exactly solvable class of models we will hereafter consider in larger detail the particular cases $f\left(
R\right) =\alpha R^{n}$ ($\alpha \neq 0$; $n \neq 1$) for matter-free universe with vanishing spatial curvature ($K=0$), since these cases are in fact analytically solvable. 
Nothing would prevent a qualitative study of more general cases. We will specify these conditions step by step in the following, when explicitly necessary. \\
In the  case $F(R)=R$ the master equation gives:
\begin{equation}
R+\tau =\beta ^{2}\ f(R)^{-2}\left( f^{\prime }\left( R\right) R-f\left(
R\right) \right) a^{-6}
\end{equation}
which simply becomes an (analytic) equation for $R$:
\begin{equation}
R+\left( 3w-1\right) \eta a^{-3\left( 1+w\right) }=\beta ^{2}\frac{%
f^{\prime }\left( R\right) R-f\left( R\right) }{\ f(R)^{2}a^{6}}
\end{equation}
In the case of polynomial (or more generally analytic, if solvable)
functions $f(R)$, solutions  $R$ of the above equation are, in
general, rational (analytic) functions of the variable $a$.
Consequently, as already remarked before, $R$ can be simply
expressed as a function of time. For any solution $R=R\left( a (t)
\right) $ one can further calculate:
\begin{equation} \label{bandc}
\begin{cases}
b=1+f^{\prime }\left( R\right) L_{d}=1+\beta ^{2}f^{\prime }\left(
R\right) \ f(R)^{-2}a^{-6} \cr
c=\frac{1}{2}\left( R+f\left( R\right) L_{d}\right) =\frac{1}{2}\left(
R+\beta ^{2}\ f(R)^{-1}a^{-6}\right)
\end{cases}
\end{equation}
The r.h.s. of the generalized FRW equation becomes:
\begin{equation}
\left( \frac{\overset{.}{a}}{a}+\frac{\overset{.}{b}}{2b}\right)
^{2}+\frac{K%
}{a^{2}}=\frac{\beta ^{2}}{6b\ f(R)a^{6}}\ +\frac{c}{3b}+\frac{(3w+1)\eta}{%
6ba^{3\left( 1+w\right) }}
\end{equation}
and it is  consequently a rational function of $a$.\\
Let us now consider  the specific example where  $f\left(
R\right) =\alpha R^{n}$ ($\alpha \neq 0$; $n \neq 1$). It is supposed that there is no matter content
in
these models.
  We thus obtain, from the structural equation:
\begin{equation}
R=\alpha R^{n}\left( n-1\right) L_{d}
\end{equation}
so that we have  $\alpha R^{n}L_{d} \equiv f\left( R\right)
L_{d}=\frac{1}{n-1}R$. Apart from the trivial solution
$R=0$ (which implies $\ b=1, \; c=0$  and $\left(
\frac{\overset{.}{a}}{a}
\right)
^{2}+\frac{K}{a^{2}}=0$) there exists also a nontrivial solution
corresponding
to the case $1=\alpha
R^{n-1}\left( n-1\right) L_{d}$. In this non-trivial case  the
time-depending parameters are found to be ($n\neq \frac{1}{2}$ to ensure $b \neq 0$):
\begin{equation}
\begin{cases}
b=1+f^{\prime }\left( R\right)
L_{d}=1+\frac{n}{n-1}=\frac{2n-1}{n-1}\cr
c=\frac{1}{2}\left( R+\alpha R^{n}L_{d}\right) =\frac{1}{2}
R\left( 1+\frac{1}{n-1}\right) =R\frac{n}{2\left( n-1\right) }
\end{cases}
\end{equation}
The conformal parameter $b$ is constant and the modified Friedmann
equations
are
\begin{equation}
\left( \frac{\overset{.}{a}}{a}\right) ^{2}+\frac{K}{a^{2}}=\frac{1}{6}\
\frac{\alpha R^{n}L_{d}}{b}+\frac{c}{3b}=\frac{\alpha
R^{n}L_{d}}{6b}+\frac{n%
}{6\left( 2n-1\right) }R=\frac{1}{6}R\left( \frac{1}{\left( n-1\right)
b}+%
\frac{n}{2n-1}\right)
\end{equation}
which in turn  can be re-written in the simplified form:
\begin{equation}
\left( \frac{\overset{.}{a}}{a}\right) ^{2}+\frac{K}{a^{2}}=\frac{1}{6}R%
\frac{n+1}{2n-1}
\end{equation}
As we already stressed before, the coefficient $b$ is now constant; then
$R=g^{\mu \nu }R_{\mu \nu
}\left( bg\right) =R\left( g\right) $ is the true Ricci scalar of the
FRW
metric $g$, which is known to be equal
\begin{equation}
R\left( g\right) =6\left[ \frac{\overset{..}{a}}{a}+\left(
\frac{\overset{.}{%
a}}{a}\right) ^{2}\right] =6H^{2}\left( 1-q\right)
\end{equation}
where $q$ is the {\it deceleration parameter}. Hence the
  two metrics $g$ and $h$ are equivalent, apart from a constant
conformal factor. The generalized Friedmann equation (specified now to
the exactly solvable case $K=0$) takes the form:
\begin{equation}
H^{2}=H^{2}\frac{n+1}{2n-1}\left( 1-q\right)
\end{equation}
which immediately gives the value of the deceleration parameter
\begin{equation}\label{q1}
q=\frac{2-n}{1+n}=-1+\frac{3}{n+1}\, 
\end{equation}
without having to solve explicitly an  ordinary differential equation for the scale factor $a(t)$. \\
It is however possible to obtain the explicit dependence of the scale factor on time. We recall that it has 
been assumed from the beginning (see (\ref{effLdll})) that $V(\phi)=0$ and consequently the field equation 
for the scalar field  is:
\begin{equation}
\frac{d}{dt}(\sqrt{g}R^{n}\overset{.}{\phi })=0
\end{equation}
which in turn implies that  $\sqrt{g}R^{n} \overset{.}{\phi }=const$ or
equivalently  that on-shell  $R^{2n}L_{d}=\beta ^{2}a^{-6}$.
From the structural equation (\ref{bandc}) we consequently obtain, in the case under discussion, that $R=\ \left[ \left( n-1\right) \alpha
\beta^{2}\right] ^{\frac{1}{n+1} }a^{-\frac{6}{n+1}}$. Therefore, again under the explicit further assumption  $K=0$, we  have:
\begin{equation}\label{jyt}
\overset{.}{a}=\left[ \frac{n+1}{6\left( 2n-1\right) }\right]
^{\frac{1}{2}}
\left[ \left( n-1\right) \alpha \beta ^{2}\right] ^{\frac{1}{2\left(
n+1\right) }}a^{1-\frac{3}{n+1}}=L(n, \alpha, \beta) a^{1-\frac{3}{n+1}}
\end{equation}
where we have defined for simplicity the constant factor $L(n, \alpha, \beta) =\left[ \frac{n+1}{6\left( 2n-1\right) }\right]
^{\frac{1}{2}} \left[ \left( n-1\right) \alpha \beta ^{2}\right] ^{\frac{1}{2\left( n+1\right) }}$. Solving this 
simple differential equation we get the explicit expression of the scale factor as a function of time:
\begin{equation}
a(t)= \left[  \frac{3 L (n, \alpha, \beta)}{n+1}  \right]^\frac{n+1}{3}(t_s\pm t)^{\frac{n+1}{3}}
\end{equation}
where $t_s$ is some integration constant. 
More precisely, Friedmann equation in the simplest form $H^2\sim a^{-2\gamma}$ (or equivalently $\overset{.}{a}\sim a^{1-\gamma }$), where $\gamma =\frac{ 3\left( w_{eff}+1\right) }{2}$, implies that  the deceleration parameter is $\ q=\gamma -1$. Moreover, in the case $\gamma>0$, choosing the solution corresponding to an expanding universe one has
$a\sim t^{\frac{1}{\gamma}}$, i.e. a forever expanding universe with Big Bang at the origin $t_s=0$. Conversely in the case $\gamma<0$, changing the arrow of time in order to avoid shrinking solutions (see \cite{GADED}), we obtain that $a\sim
{(t_s-t)}^{\frac{1}{\gamma}}$, i.e. we find a universe with a final Big Rip singularity. Thus 
$H\sim (t_s-t)^{-1}\rightarrow\infty$ at finite future time $t_s$.\\
Coming back to the case under consideration governed by eqn. (\ref{jyt}) it follows that the effective value of $w$ (for $n\neq -1$) is
\begin{equation} \label{gert}
w_{eff}=\frac{1-n}{n+1}=-1+\frac{2}{n+1}
\end{equation}
The reliability conditions one has to
assume are respectively $\frac{n+1}{2n-1}>0$ and $\left(
n-1\right)\alpha >0$, which follow directly from (\ref{jyt}). Models for
accelerating universe occur
  consequently for $n>2$ or $n<-1$. The first case corresponds to an effective
quintessence when $-1<w_{eff}<-\frac{1}{3}$  and one deals with an
initial Big Bang type singularity at the origin. In contrast, the
second case leads to an effective phantom
($w_{eff}<-1$) (without its explicit introduction) with a Big Rip type
final
singularity (for a discussion of it in modified gravity see \cite{phantom}),
with the asymptotic behavior $a(t_s) \sim + \; \infty$.\\
The case $n=-1\ $\ should be treated separately. In this case $\overset{.}{a }=0$; this implies that the trivial solution $R=0=\overset{.}{\phi }$ emerges
 and we obtain  a non-expanding universe model with $a=a_{0}=constant$.\\
\subsubsection{The radiating universe}
One can  also combine the above case $F(R)=R$ with the presence of radiation.  Adding
radiating matter ($w=\frac{1}{3}$) gives rise to the following
modified Friedmann equation:
\begin{equation}
\left( \frac{\overset{.}{a}}{a}+\frac{\overset{.}{b}}{2b}\right)
^{2}+\frac{K%
}{a^{2}}=\frac{\beta ^{2}}{6b\ f(R)a^{6}}\
+\frac{c}{3b}+\frac{\eta}{3ba^{4}}
\end{equation}
To obtain exactly solvable models it is again convenient to specify to the case $f(R)= \alpha R^n$; we thus obtain ($n\neq
-1,\frac{1}{2}, 1$):
\begin{equation} \label{oiuy}
H^{2}=\frac{n+1}{6\left( 2n-1\right) }\left[ \left( n-1\right) \alpha
\beta
^{2}\right] ^{\frac{1}{n+1}}a^{-\frac{6}{n+1}}+\frac{(n-1)\eta}{3\left(
2n-1\right) }a^{-4}
\end{equation}
and we are able to compute the deceleration parameter to be
\begin{equation}
q=\frac{\frac{2-n}{3\left( 2n-1\right) }\left[ \left( n-1\right) \alpha
\beta ^{2}\right] ^{\frac{1}{n+1}}a^{-\frac{6}{n+1}}+\frac{2(n-1)}{3\left(
2n-1\right) }\eta a^{-4}}{2H^{2}}
\end{equation}
A detailed analysis of this  last equation is more complicated but
the final answer is similar to the previous case:
accelerated universe can be obtained only for $n>2$ or
$n<-1$.\footnote{Notice that some cases are excluded by the
condition $H^2>0$ and others give rise to deceleration.} We stress,
moreover, that the asymptotic value of the deceleration parameter for big enough
radius $a(t)$ in both cases remains the same, i.e.
\[
q_{asymp}=\frac{2-n}{n+1}
\]
as in the radiation-free example (\ref{q1}).
Because of this the singularity types should be
the same as well. For $n<-1$ (which, by the way, corresponds to the case
$\gamma<0$), besides the final Big Rip singularity
one can encounter the initial Big Bang singularity caused by
radiation. To be more precise this follows from the qualitative behavior
of $H^2$ (see (\ref{oiuy})):
\[
H^{2} \sim A a^{-2 \gamma}+ B a^{-4} \qquad \quad (n<-1)
\]
where $A$ and $B$ can be simply deduced from  (\ref{oiuy}). In the case
$\gamma<0$ the first term dominates in late time universe, giving
rise to a Big Rip like singularity. On the contrary, in the same case
$\gamma<0$, we get that small values of $a$ correspond to a behavior of
$H^{2} \sim  B a^{-4}$, which implies the presence of
a Big Bang singularity at early time universe. In the case $n>2$
($\gamma>0$) one obtains an accelerated and expanding model, 
such that the initial Big Bang is still present.  Finally we remark that for
$n>2$ there exists a critical value for the radius
\begin{equation} \label{topo}
a_{crit}^{2}=\left[\frac{2(n-1)}{n-2}\eta\right] ^{\frac{n+1}{2n-1}}
\left[ \left( n-1\right) \alpha
\beta ^{2}\right] ^{-\frac{1}{2n-1}}
\end{equation}
which provides a transition from deceleration to acceleration.\\
To summarize we have thus proven that, also in the Palatini formalism,
non-linear gravity-matter systems based on the Lagrangian  (\ref{lagruno}) with $F(R)=R$ 
lead to accelerating universe in the same way as it happens in the
metric formalism. Big Rip singularities appear in the case of effective phantom
models in vacuum universe, while quintessence models contain Big Bang like
singularities. We have moreover shown that in the case of radiation
models, where a radiation-like fluid is additionally
considered, we have both Big Rip and Big Bang singularities appearing.\\
\\
A full set of possible FRW cosmological solutions of the above non-linear gravity-matter system based  on the Lagrangians  (\ref{lagruno}) with $F(R)=R$ has been found,  under the specific hypotheses $f(R) \sim R^n$, $V(\phi)=0$ and $K=0$ both in the case of vacuum and radiating universes.  It turns out that the same solutions were already found and studied, for the same class of Lagrangians, in the framework of the purely metric formalism \cite{GADED} where they however represent  a proper subset in  the set of all possible solutions. We think that a bit of discussion is here appropriate on these arguments.\\
It is well known that, in the matter free case $L_d=0$, $L_{mat}=0$, i.e.  for pure $F(R)$-gravity, solutions of the Palatini formalism represent in general a subset in the set of all possible solutions allowed by the purely metric formalism; see \cite{magna}. As a matter of fact in that case the conformal factor $b=F^\prime(R)$ relating the two metrics $g_{\mu \nu}$ and $h_{\mu \nu}$ becomes constant by virtue of the master equation; this implies that the scalar $R$ is equal to the Ricci scalar of the metric $g$. As far as the field equations in the metric formalism are concerned, it implies also that the second order differential operator $\nabla_\mu\nabla_\nu -g_{\mu\nu}\square$ acting on $F^\prime(R)$ vanish identically; see e.g.  \cite{magna}. This finally implies in turn that the field equations in the metric formalism, expressed in the metric $g_{\mu \nu}$,  almost reproduce  the field equations of the Palatini formalism.\\
It is easy to check that exactly the same mechanism acts in our case, provided that $f(R) \sim R^n$. As a consequence we obtain that, in the particular example we have here analyzed (i.e. $F(R)=R$), the Palatini formalism provides at the same time solutions to  the metric one.


\subsection{ First-order scalar $F\left( R\right) $-$f\left( R\right) $
non-minimal coupling}
We generalize hereafter the analysis to the more general family of models already introduced in (\ref{lagruno}) for  non-linear gravity-scalar system. To obtain exactly solvable models, which are preferable as they allow us to discuss relevant physical consequences, we will specify step by step the necessary conditions ( $f\left(
R\right) =\alpha R^{n}$ ($\alpha \neq 0$; $n \neq 1$), $\tau=0$ and $K=0$).  We thus consider the more general class of Lagrangians:
\begin{equation} \label{lagruno4}
L=\sqrt{g}\left( F(R)+f\left( R\right) L_{d}\right) +L_{mat}
\end{equation}
where  $L_{d}=-\frac{1}{2}g^{\mu \nu }\partial _{\mu }\phi \partial_{\nu }\phi +V\left( \phi \right)$ and  $L_{mat}$  represents a matter Lagrangian but now $F(R) \neq R$.  We get now that the r.h.s. of the generalized Friedmann equation (\ref{utrec}) becomes:
\begin{equation} \label{jhdecvb}
\left( \frac{\overset{.}{a}}{a}+\frac{\overset{.}{b}}{2b}\right) ^{2}+\frac{K
}{a^{2}}=\frac{\beta ^{2}}{6b\ f(R)a^{6}}\ +\frac{c}{3b}+\frac{(3w+1)\eta }{
6ba^{3\left( 1+w\right) }}
\end{equation}
We proceed by specifying the model to the concrete particular case ($n\neq 1,$ $\ m\neq 2$)
\begin{equation}
\begin{cases}
f\left( R\right) =\alpha R^{n}\cr
F\left( R\right) =\omega R^{m}
\end{cases}
\end{equation}
without matter (i.e. $\tau=0$ and $\tau_{\mu \nu}^{mat}=0$) which leads again to an exactly solvable class of models. Combining the master equation
with the on-shell value $f(R)^{2}L_{d}=\beta ^{2}a^{-6}$ (see (\ref{effLd})) it is simply possible to obtain the Ricci scalar as a function of the radius $a$:
\begin{equation}
R=\xi a^{-\frac{6}{m+n}}
\end{equation}
where the coefficient $\xi$ is given by $\xi =\left[ \frac{\beta^2 \left( n-1\right) }{\omega \left( 2-m\right) }
\right] ^{\frac{1}{n+m}}$ . In this case it follows that:
\begin{equation}
\begin{cases}
b =\left( m\omega \xi
^{m-1}+n \alpha^{-1} \beta^2 \xi ^{-\left( n+1\right) }\right) a^{-\frac{6\left(
m-1\right) }{m+n}}=b_{0}a^{-\frac{6\left( m-1\right) }{m+n}}\cr
c=\frac{1}{2}
\left( \omega \xi ^{m}+\alpha \beta \xi ^{-n}\right) a^{-\frac{6m}{m+n}
}=c_{0}a^{-\frac{6m}{m+n}}
\end{cases}
\end{equation}
Since now $b=b_{0}a^{-\lambda }$ is proportional to a power of $a$, it follows that the
generalized Hubble parameter  $\hat{H}=\frac{\overset{.}{a}}{a}+\frac{\overset{.}{b}}{2b}$ is, up to the constant
factor   $1-\frac{3(m-1)}{m+n}$, proportional to $H$. Therefore one has (again in the solvable case $K=0$):
\begin{equation}
H^{2}\sim \lbrack \frac{\beta ^{2}}{6b_{0}\alpha \xi ^{n}}\ +\frac{c_{0}}{
3b_{0}}]a^{-\frac{6}{m+n}}\sim a^{-\frac{6}{m+n}}
\end{equation}
This further implies that \ $\overset{.}{a}\sim a^{1-\frac{3}{m+n}}$ (which resembles
 the  case $m=1$, already treated in Section \ref{sec3}). Consequently:
\[
a(t) \sim (t_s\pm t)^{\frac{m+n}{3}}
\]
The deceleration parameter turns out to be $q=-1+\frac{3}{n+m}$. As for the very particular case, already considered in Section \ref{sec3}, we obtain by a similar analysis an expanding universe in the case $m+n>3$, presenting a Big Bang like singularity; conversely, for $m+n<0$ the  expanding solution leads to a Big Rip finite time singularity.\\
We stress that now the Palatini formalism and the metric one are completely different both from a qualitative and from a quantitative viewpoint. 
This is related to the fact that in this case the conformal factor $b$ as given by (\ref{netnet}) is no longer constant.
Moreover we have seen that in the Palatini formalism field equations are easily solvable also in this more general case, while in the metric formalism this is by no means true.


\subsubsection{The case $F(R)=R+\mu G\left( R\right) $}
\noindent We can consider the interesting case when  $F(R)=R+\mu G\left( R\right) $, which reproduces the case already  treated in subsection \ref{trippa} in the case $\mu=0$ (or in the case $G(R)=0$). The corresponding Lagrangian is then:
\begin{equation}
L=\sqrt{g}\left( R+\mu G(R)+f\left( R\right) L_{d}\right)
+L_{mat} (\Psi)
\end{equation}
and the master equation is:
\begin{equation} \label{matd}
R+2\mu G \left( R\right) -\mu G^{\prime }\left( R\right) R+\tau =\left(
f^{\prime }\left( R\right) R-f\left( R\right) \right) L_{d}-f\left(
R\right) V\left( \phi \right)
\end{equation}
We specialize moreover to the case $f\left( R\right) =\alpha R^{n},$ \ \ $G \left(
R\right) = R^{2n}$,  $V(\phi)=0$ and for simplicity $L_{mat}=0$ (the case of vacuum universe). Equation(\ref{netnet}) gives in this case:
\begin{equation}
\begin{cases}
b=1+f^{\prime }\left( R\right) L_{d}+\mu G^{\prime }\left( R\right) =1+
\frac{n}{n-1}=\frac{2n-1}{n-1} \cr
c=\frac{1}{2}\left( R+f\left( R\right) L_{d}+\mu G\left( R\right) \right) =R
\frac{n}{2\left( n-1\right) }-\frac{1}{2} \mu G\left( R\right)
\end{cases}
\end{equation}
where the value of the conformal factor $b$ is again constant, as much as $f^{\prime }\left( R\right) L_{d}+ \mu G^{\prime }\left( R\right)
=b-1$ is constant too. This implies that the solutions of the Palatini formalism
are at the same time solutions of the
metric formalism, as much as in the case $F(R)=R$ previously examined (see our discussion after formula (\ref{topo})).
It is however clear  that in this case  the Palatini formalism 
is much easier  to handle and more effective
then the purely metric formalism (even if it provides the same 
solutions). Substituting the
actual value of $L_{d}$ along solutions  into the master equation, 
 as given by (\ref{effLd}), we obtain now:
\begin{equation}
R^{n+1}+2\mu (1-n)R^{3n}=\alpha \left( n-1\right) \beta ^{2}a^{-6}
\end{equation}
Since $R$ is positive, in order to get an approximate solution of the polynomial equation above, one needs to distinguish two cases:
\begin{enumerate}
\item Small values of $R$ ($R<<1$) and $n>>0$ or $R>>1$ and $n<<0$ lead to
\[
R\sim a^{-\frac{6}{n+1}}
\]
\item  Big values of $R>>1$  and $n>>0$ or $R<<1$ and $n<<0$ lead instead to
\begin{equation} \label{maui}
R\sim a^{-\frac{2}{n}}
\end{equation}
\end{enumerate}
In any case we obtain  a power law $R\sim a^{-\lambda }$, where $\lambda = \frac{6}{n+1}$ in the first case and $\lambda = \frac{2}{n}$ in the second one. \\
Owing to the fact that $b$ is constant it is now easy to obtain the modified Hubble equations from (\ref{jhdecvb}) (in that case $g$ is obviously chosen to be a FRW metric for cosmological applications):
\begin{equation}
H^{2}\sim -Ka^{-2}+\frac{\beta ^{2}\left( n-1\right) }{6\left( 2n-1\right) }
R^{-n}a^{-6}\ +\frac{n}{6\left( 2n-1\right) }R-\frac{\mu \left(
n-1\right) }{6\left( 2n-1\right) }R^{2n}
\end{equation}
which further  re-writes as:
\begin{equation}
H^{2}\sim -Ka^{-2}+Aa^{-(6-n\lambda )}\ +Ba^{-\lambda }+Ca^{-2n\lambda}
\end{equation}
with suitable chosen constants $A, B, C, D$.
In the first case, recalling that $\lambda =\frac{6}{n+1}$, one  has:
\[
H^{2}\sim -Ka^{-2}+(A+B)a^{-\frac{6}{n+1}}+Ca^{-\frac{12n}{n+1}}
\]
For late universe ($a>>1$) the term $a^{-\frac{6}{n+1}}$ dominates and the deceleration parameter is $q=\frac{2-n}{n+1}$. This gives the same qualitative behavior already found in the case $F(R)=R$ previously discussed. For early universe ($a<<1$), instead, the term $a^{-\frac{12n}{n+1}}$ dominates,   
providing a Big Bang initial singularity.\\
Correspondingly, for $R\sim a^{-\frac{2}{n}}$ the Friedmann equation reduces to
\[
H^{2}\sim -Ka^{-2}+Ba^{-\frac{2}{n}}+(A+C)a^{-4}  
\]
and consequently the term $a^{-\frac{2}{n}}$ dominates in late universe.
However this  contradicts to our initial assumptions $R>>1$  and $n>>0$ or $R<<1$ and $n<<0$ for that case; see eqn. (\ref{maui}).
For early universe, conversely, the term $a^{-4}$ dominates and we obtain a universe fulfilled by radiation with a Big Bang initial singularity. 


\section{ The cosmological constant problem}
One of the alternative proposals to explain the present acceleration of
the universe is related to dynamical cosmological constant models. This
approach, however, has a lot of  unsolved issues related to the fact that there is
no known mechanism, up to now, that guarantees zero or nearly zero energy in
a stable energy configuration for the universe.
Using cosmological constant-like models, we have to  face with some
problems concerning: (i) the small amount of energy of the vacuum, which is much
smaller than we estimate it to be (the so-called \textit{cosmological constant
problem}); (ii) the nature of the dark energy which seems to dominate
the
universe and; (iii)  the \textit{coincidence problem} between the actual
density of dark energy in the universe and the actual matter density
\cite{carrol2}. \\
It has been recently proposed in \cite{muko} that the apparent value of
the cosmological constant can be alternatively  determined 
  by dynamical considerations in the framework of alternative theories of gravity. 
The true value of the vacuum energy is not zero,
  but settles down to a nearly zero value when the curvature of the universe
  reaches nearly zero values.  This mechanism is produced by means of a
feedback process in which the scalar equation depends on curvature,
  in such a way that the divergent coefficient of the kinetic term of
  the scalar field forbids  the lowest energy state to be reached.
  The potential of the scalar field is chosen to have a minimum
at a generic negative value for the vacuum energy. However,
  the singular kinetic term of the scalar field makes the dynamics
of the scalar field  to stop at zero, where the field itself gets frozen.
(Of course, some fine-tuning of dynamical Lagrangian is assumed).\\
This theory has been proven to have some very important properties,
  i.e. it is stable under radiative corrections and it leads to stable
dynamics.
  No fine-tuning of the potential is necessary in this case, neither
the introduction of an anthropic principle. \\
The phenomenological theory introduced in \cite{muko} slightly differs
from the
one we studied in (\ref{lagruno}). The Lagrangian density is:
\begin{equation} \label{lagrduo}
  L=\sqrt{g}\left( \frac{R}{2 \kappa^2}+ \alpha R^2+  f \left( R \right)
L_{kin}
- V (\phi) \right)
\end{equation}
where $L_{kin}= -\frac{1}{2}g^{\mu \nu }\partial _{\mu }\phi $
$\partial_{\nu
}\phi $ is the kinetic term of a scalar field Lagrangian and f(R) is a
function
of the Ricci scalar, which is assumed to be divergent at $R=0$. It
should also
be noticed that (\ref{lagrduo}) coincides with (\ref{lagruno}) for
$\alpha=0$, $F(R)=R$  
and $V(\phi)=0$. \\
We want here to examine if the same mechanism, generating a dynamical
cosmological constant with suitable properties, works in the first order
formalism. Let $g$  be a flat FRW metric (\ref{FRW}). First of all, even
in the
Palatini formalism the same field equations hold for the scalar, i.e. they
have
the same form as ones in the metric formalism \cite{muko}:
  \begin{equation}
\frac{d}{dt} \left[ \dot{\phi} f(R) \right]+ 3 H \dot{\phi} f(R)+V^\prime (\phi)=0
\end{equation}
As far as field equations for the metric field $g$ are concerned, we obtain in
analogy
with (\ref{eqng}) that:
\begin{equation}   \label{eqn33}
\left( \frac{\sqrt{g}}{2 \kappa^2}+ \sqrt{g} \alpha R \right) \left(
R_{\mu \nu
}-\frac{1}{2}R g_{\mu \nu } \right) +f^{\prime }\left( R \right)
L_{kin} R_{\mu \nu }-\frac{1}{2} f \left( R \right) L_{kin} g_{\mu \nu }-
{1
\over 2} V(\phi) g_{\mu \nu }=0
\end{equation}
and the scalar-valued structural field equation, controlling the solutions
of
(\ref{eqn33}), can be simply obtained by taking the trace of
(\ref{eqn33}) itself.
This implies that,
considering $\phi$ to evolve slowly, we have that $V(\phi)$ can be
approximated
as a linear function, see \cite{muko}:
\begin{equation}   \label{eqn34}
V(\phi) \simeq c \kappa^{-3} (\phi-\phi_0)
\end{equation}
and the asymptotic behavior of $\phi$ is consequently:
\begin{equation}
\kappa^2 \dot{\phi} f(R) \sim - c \kappa^{-1} H^{-1}
\end{equation}
This expression follows from the consideration that the term $\frac{\dot
H}{H^2}  $ is nearly zero at small energies. From field equations  (\ref{eqn33}),
in the approximation that the kinetic term is small compared to the
potential
term and assuming $\frac{\dot H}{H^2} \simeq 0$, one obtains in analogy
with
\cite{muko}:
\begin{equation}
3 H^2 \simeq \kappa^2 V
\end{equation}
This is exactly the same result already obtained in \cite{muko} in the
metric
formalism, proving that field equations  from the Lagrangian
(\ref{lagrduo})
have the same limit at low energies, both in the Palatini and the metric
formalism. It follows (see  \cite{muko} for details), that at low energies
$\phi$
stalls at nearly $V=0$, where $V$ includes all the contributions to
the
cosmological constant. This consequently implies that the value of the
cosmological constant stalls at a very small but non-zero value, in the
limit
of small curvatures of the metric $g$, i.e. at small energies for the
gravitational field. It has also been proven in  \cite{muko} that this
behavior
is stable under the effect of radiative corrections and leads to stable
dynamics, provided that the minimum of the potential always occurs at a
negative value. Hence, the same dynamical mechanism to solve the
cosmological
constant problem works in Palatini formulation.

\section{ Conclusions}

In summary, the Palatini formulation of non-linear gravity-matter system
has been developed. Using a scalar field Lagrangian as matter it is shown
that such a theory provides the effective quintessence or effective
phantom cosmology at late times in the same qualitative way as
in the metric version. In addition, the gravity assisted dark energy
dominance mechanism also occurs. It is shown that an account of radiation
maybe
done with possible emergence of accelerating cosmology again.
It is interesting to notice that in this case the emerging cosmology
may contain Big Bang at early times and Big Rip at late times (for
general
review of late-time singularities, see ref \cite{shinji} and references
therein). Eventually, the account of quantum effects near to singularities
is necessary to see if the singularities are realistic or not.
It would be really interesting to perform such a study within
the Palatini approach.

Any Palatini model of gravity possess its purely metric counterpart which leads to 
fourth order field equations. An advantage of Palatini formulation rely on second
order field equations which turns out to be more easy to solve. In some very 
specific situations both approaches give rise to the same solutions but in 
general it is not true. The same occured  here in the case of cosmological 
application.  Although, one can expect that metric formulation may also produce 
accelerated  solutions  thus, however, it is much more difficult (if possible at all) 
to find them explicitly out. In this sense the Palatini formalism is more easy to handle and simpler
to analyse then the corresponding metric formalism.

It is obvious that any resonable model of gravity should satisfy the standard 
solar system tests. In the context of modify gravity this problem has been studied 
recently by several authors \cite{AFRT} (both in the metric and Palatini approaches). 
It has been shown that, in principle, Palatini formalism  provides 
a good Newtomian approximation. Particularly, the Schwartzschild limit of the theory 
can be recovered for positive powers $n$ of the scalar $R$, greater than one $n \geq 1$. In the case of negative 
powers one should expect de-Sitter or anti-de-Sitter black holes instead \cite{BNOV}.
 
It has been also explicitly demostrated that dynamical mechanism to resolve
the cosmological constant problem suggested in metric formulation
works also in Palatini formulation
(within the same choice of scalar-tensor Lagrangian). Again, it is very
interesting to understand the role of quantum effects (as this is exactly the region
where
they are expected to be essential) in the above mechanism.
However, to do so one needs to evaluate quantum effects in the Palatini
formulation, something that is rather far from being a trivial task.

Finally, in this paper we have limited ourselves to the consideration
of non-linear scalar-gravity theories in the Palatini formulation.
However, it is clear that non-linear spinor-gravity system (where the spinor
part can be constructed from NJL-like model coupled to gravity \cite{muta})
maybe more interesting from a cosmological point of view. Indeed, unlike
 the covariant derivative of  a scalar which is nothing but a partial derivative, 
the covariant derivative of a spinor includes instead  the
connection, what definitely leads to new features in the spinor analog of Eq.7.
This will be discussed elsewhere.

\section{Acknowledgments}
All the authors acknowledge the referee of this paper for suggestion to write the
Appendix. Two of us (G.A and A.B) are grateful to Prof. A. P\c{e}kalski for
providing us in an excellent working conditions during 41st Winter
School of Theoretical Physics in L\c{a}dek. A.B. is supported by KBN
grant 1P03B01828. G.A. is supported by the I.N.d.A.M. grant: "Assegno di collaborazione ad attivit\'a di ricerca a.a. 2002-2003".
 M.F. and G.A. are also supported by MIUR: PRIN 2003 on
``\emph{Conservation laws and thermodynamics in continuum mechanics and field theories}'', and by INFN ({ \it Iniziativa specifica NA12}).

\appendix
\section{Some useful relations in the Palatini formalism} 
We include hereafter some useful relation in the Palatini
formalism for the benefit of the general reader. The Palatini 
formalism can be considered as a metric-affine theory for relativistic theories.
Metric and connection are firstly considered as independent objects. We remark however 
that, without needing any reference to a metric, a connection can be introduced from
which one can construct covariant derivatives. In order to produce
covariant derivatives which transform as true tensors, under a
coordinate transformation the connection itself needs to transform as:
$$\Gamma^{\lambda}_{\mu\nu}\rightarrow
{\partial x^{\prime\lambda} \over \partial x^{\rho}}
{\partial x^{\tau} \over\partial x^{\prime \mu}}
{\partial x^{\sigma} \over\partial x^{\prime
\nu}}\Gamma^{\rho}_{\tau\sigma}+
{\partial x^{\prime\lambda} \over\partial x^{\rho}}
{\partial^2 x^{\rho} \over\partial x^{\prime \mu}\partial x^{\prime
\nu}}~~.
$$
Then for such a connection the quantity
$$R^{\lambda}_{\mu\nu\kappa}={\partial \Gamma^{\lambda}_{\mu\nu} \over
\partial x^{\kappa}}
-{\partial \Gamma^{\lambda}_{\mu\kappa} \over
\partial x^{\nu}}
+\Gamma^{\eta}_{\mu\nu} \Gamma^{\lambda}_{\kappa\eta}
-\Gamma^{\eta}_{\mu\kappa} \Gamma^{\lambda}_{\nu\eta} $$
will then be a true tensor, regardless of whether the connection is or
is not related to the metric in the standard metric based Christoffel
symbol way
$$ \Gamma^{\lambda}_{\mu\nu}(g)={1 \over 2}g^{\mu\sigma}[
\partial_{\mu}g_{\sigma\nu}
+\partial_{\nu}g_{\mu\sigma}-\partial_{\sigma}g_{\mu\nu}]~~.$$
However, noting that the metric based $ \Gamma^{\lambda}_{\mu\nu}(g)$ also
transforms as
$$\Gamma^{\lambda}_{\mu\nu}(g)\rightarrow
{\partial x^{\prime\lambda} \over \partial x^{\rho}}
{\partial x^{\tau} \over\partial x^{\prime \mu}}
{\partial x^{\sigma} \over\partial x^{\prime
\nu}}\Gamma^{\rho}_{\tau\sigma}(g)+
{\partial x^{\prime\lambda} \over\partial x^{\rho}}
{\partial^2 x^{\rho} \over\partial x^{\prime \mu}\partial x^{\prime
\nu}}~~,
$$
we see that the difference between the general connection and the
Christoffel symbol connection transforms as
$$\Gamma^{\lambda}_{\mu\nu}-\Gamma^{\lambda}_{\mu\nu}(g)\rightarrow
{\partial x^{\prime\lambda} \over \partial x^{\rho}}
{\partial x^{\tau} \over\partial x^{\prime \mu}}
{\partial x^{\sigma} \over\partial x^{\prime
\nu}}[\Gamma^{\rho}_{\tau\sigma}-\Gamma^{\rho}_{\tau\sigma}(g)]~~,
$$
with the difference thus being a true tensor. Then, with covariant
derivatives being linear in the connection, the difference between the
quantity $T^{\mu\nu}_{;\mu}$ (where $T^{\mu\nu}$ is the stress-energy tensor of matter), 
as calculated with $\Gamma^{\rho}_{\tau\sigma}$, and
$T^{\mu\nu}_{;\mu}$, as calculated with $\Gamma^{\rho}_{\tau\sigma}(g)$, is
itself a true tensor, and since $T^{\mu\nu}_{;\mu}$ as calculated with
$\Gamma^{\rho}_{\tau\sigma}$ is necessarily a true tensor, it follows that
$T^{\mu\nu}_{;\mu}$ as calculated with $\Gamma^{\rho}_{\tau\sigma}(g)$ is
one as well. It is consequently meaningful to take the covariant derivative of 
equation of motion (\ref{eqng})  with respect to the Christoffel symbol based
connection: it transforms like a true tensor.


\begin{thebibliography}{99}


\bibitem{Perlmu} Perlmutter, S. et al. 1999, ApJ, 517, 565
(astro-ph/9812133 );
Perlmutter el al. Nature 404 (2000) 955
    ; Riess, A. G. et al. 1998, AJ, 116, 1009 (astro-ph/9805201); C.L.
Bennett
et al.,  Astrophys.J.Suppl. 148 (2003) 1 (astro-ph/0302207); C.B.
Netterfield
et al., Astrophys.J. 571 (2002) 604-614 (astro-ph/0104460); N.W. Halverson
et
al., Ap.J. 568, 38 (2002), (astro-ph/0104489);
  A.H. Jaffe et al., Phys. Rev. Lett. 86, 3475 (2001), (astro-ph/0007333);
A.E.
Lange et al., Phys. Rev. D 63, 042001 (2001); A. Melchiorri et al., Ap.J.
563,
L63 (2000), (astro-ph/9911445); Spergel, D. N. et al. 2003, ApJS, 148,
175;
Verde, L. et al. 2002, MNRAS, 335, 432.

\bibitem{GADED} S. Nojiri and S. Odintsov, Phys.Lett. {\bf B599} (2004)
137
(astro-ph/0403622).

  \bibitem{muko} S. Mukohyama and L. Randall, hep-th/0306108.

\bibitem{dolgov} A.D. Dolgov and M. Kawasaki, astro-ph/0307442.

\bibitem{PBM} V. Pettorino, C. Baccigalupi and G. Magnano,
astro-ph/0412334.

\bibitem{carrol2} S.M. Carroll, V. Duvvuri, M. Trodden and M.S. Turner,
(astro-ph/0306438); S. Capozziello, S. Carloni and A. Troisi, "Recent
Research
Developments in Astronomy and Astrophysics" -RSP/AA/21-2003
(astro-ph/0303041);
S. Capozziello, Int. J. Mod. Phys. D11, 483 (2002); S. Capozziello, V.F.
Cardone, S. Carloni and A. Troisi, Int. J. Mod. Phys. D12, 1969 (2003).

\bibitem{eli} G. Cognola, E. Elizalde, S. Nojiri, S.D. Odintsov and
S. Zerbini, hep-th/0501096.

\bibitem{FFV} M.Ferraris, M.Francaviglia and I.Volovich, Nouvo Cim. B108
(1993)
1313 (gr-qc/9303007);
M.Ferraris, M.Francaviglia and I.Volovich, Class. Quant.Grav. 11 (1994)
1505.

\bibitem{metricfR}
S. Nojiri and S.D. Odintsov, Phys. Rev. D68, 123512 (2003)
(hep-th/0307288);
  GRG 36 (2004) 1765 (hep-th/0308176); Mod.Phys.Lett.A 19 (2004)627
(hep-th/0310045).

\bibitem{do}
   A.D. Dolgov
and M.
Kawasaki, astro-ph/0307285;
M. Soussa and R.P. Woodard, astro-ph/astro-ph/0308114.

\bibitem{carroll1} S.M. Carroll,  A. De Felice, V. Duvvuri, D. Easson, M. Trodden and M.S. Turner,
Phys. Rev. D71 (2005) 063513, (astro-ph/0410031);

\bibitem{ABFR} G. Allemandi, A. Borowiec and M. Francaviglia,  Phys.Rev. D
{\bf
70}, 043524 (2004), (hep-th/0403264).

\bibitem{palatinifR} D.N. Vollick, Phys. Rev. D68, 063510 (2003)
(astro-ph/0306630); X.H. Meng and P. Wang, Class. Quant. Grav. 20,
4949-4962
(2003) (astro-ph/0307354);  Class. Quant. Grav. 21, 951-960 (2004)
(astro-ph/0308031); (gr-qc/0411007);
  (hep-th/0309062);
A. Dominguez and D. Barraco,Phys.Rev. D70, 043505 (2004);
S. Capozziello, V.F. Cardone, and M. Francaviglia,astro-ph/0410135;
P. Wang, G. Kremer, D. Alves and X. Meng, gr-qc/0408058.

\bibitem{Ricci2} G. Allemandi, A. Borowiec and M. Francaviglia,  Phys.Rev.
D{\bf 70}, 103503 (2004), (hep-th/0407090).

\bibitem{FF04} L. Fatibene and M. Francaviglia, {\it Natural and gauge natural formalism for classical field theories. A geometric perspective including spinors and gauge theories}, \\ 
Kluwer Academic Publishers, Dordrecht, 2003. 

\bibitem{Koivisto} T. Koivisto, gr-qc/0505128.
\bibitem{phantom} M. Abdalla, S. Nojiri and S.D. Odintsov, hep-th/0409177.

\bibitem{magna}  G. Magnano,
{\it Proceedings of XI Italian Conference on General Relativity and Gravitation, Trieste}, (1994), (gr-qc/9511027).

\bibitem{shinji} S. Nojiri, S.D. Odintsov and S. Tsujikawa,
  hep-th/0501025.

\bibitem{AFRT}  X. Meng and P. Wang, Gen. Rel. Grav. 36, 1947 (2004);
A. E. Dom\'inguez and D. E. Barraco, Phys. Rev. D 70, 043505 (2004);
G. J. Olmo, gr-qc/0505101; G. Allemandi, M. Francaviglia, M.L. Ruggiero and A. Tartaglia,  
gr-qc/0506123; T.P. Sotiriou, gr-qc/0507027,  J.A.R. Cembranos. UCI-TR-2005-29, 2005, gr-qc/0507039.


\bibitem{BNOV} I. Brevik, S. Nojiri, S.D. Odintsov and L. Vanzo,
 Phys.Rev. D70 043520 (2004), hep-th/0401073

\bibitem{muta} T. Inagaki, T. Muta and S.D. Odintsov, hep-th/9711084,
Progr.Theor.Phys.Suppl. {\bf 127},93-193 (1997).

\end{thebibliography}
\end{document}